# Comment on "How to Realize Uniform Three-Dimensional Ellipsoidal Electron Bunches"

Yuelin Li (李跃林)

*Argonne National Laboratory, Argonne, IL 60439*

Via systematic numerical simulation we found that the recently propose 'Pancake' scheme [O. J. Luiten et al., Phys. Rev. Lett. **93**, 094802 (2004)] does not generate a beam desired for low emittance, even when compared with the more traditionally accepted cylindrical beam geometry.

A recent letter [1], extending the proposal in [2], detailed a scheme for realizing a uniform ellipsoidal (UE) electron beam to minimize the beam emittance in modern photoinjectors. The scheme exploits the space-charge force (SCF) driven expansion of a thin "pancake" (PC) beam. The authors support the validity of the scheme by comparing the emittance with that of a uniform cylindrical (UC) beam. At 0.1 nC, the emittances are shown at 0.4 and 1.2 mm mrad for the PC and the UC beams, respectively, in a DC accelerating field [1]. Unspecified but "equally impressive results" in an rf photoemission gun are also claimed [1].

It is known that at higher bunch charge, the PC scheme performs poorly [1, 3]. In this comment, based on more detailed simulations [4], we show that even at 0.1 nC, the PC scheme is inferior, not only to a generic UE beam, but also to a UC beam when proper optimization is performed. Physically, this is due to the necessary initial "pancake" geometry required by the desired SCF-driven evolution [2]. In practice, it leads to a larger initial transverse beam size, hence a larger cathode emittance. The UC and UE beams, in contrast, can be initialized at much smaller transverse beam sizes, thus a smaller cathode emittance.

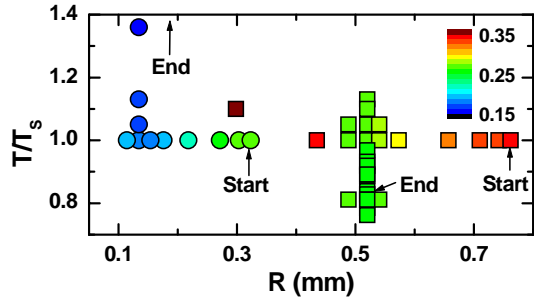

Fig. 1 Emittance at 0.2 m from the cathode as a function of bunch length $T$ and radius $R$ for a 0.1-nC PC (squares) and a UC (circles) beam in a 100-MV/m DC field, extracted from the optimization log. The starting bunch length is $T_s$= 10 ps for BC beam (full width) and 30 fs for the PC beam (FWHM). The color indicates the value of the emittance in mm mrad.

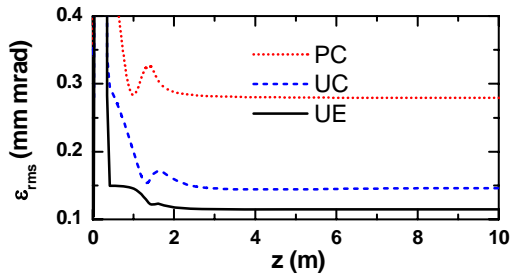

Fig. 2 Normalized emittance as a function of beam propagation distance under optimized compensation setup for 0.1-nC UC and PC beams. The optimized initial beam conditions for the UE, UC, and PC beams are: $R$=0.30, 0.30, and 0.74 mm, $T$=14, 12.6, and 0.032 ps, respectively.

This is illustrated in Fig. 1, which is the optimization log for the UC and PC beams in a DC accelerating field. It shows the emittance as a function of the laser beam radius $R$ and pulse duration $T$. The optimization for the UC (UE) beam does not converge, but the emittance reduces to 0.17 (0.09) mm mrad at $R$=0.134 (0.25) mm before the bunch lengthens beyond 10 ps. The PC beam converges at $R$=0.52 mm, which inevitably leads to a larger emittance at $\varepsilon$=0.27 mm mrad. The corresponding cathode emittances are 0.08, 0.05, and 0.17 mm mrad for the UE, UC, and PC beams, respectively.

Cathode emittance cannot be compensated, as can be shown in the more realistic rf phototinjector scenario with emittance compensation [5]. In this case, the rf-induced emittance [6] places an upper limit on the bunch length, thus the optimization converges also for UE and UC beams. The optimized emittances are $\varepsilon$=0.11, 0.16, and 0.29 mm mrad, and the corresponding cathode emittances are 0.1, 0.11, and 0.24 mm mrad for the UE, UC, and the PC beams, respectively (Fig. 2). The PC scheme is again harmed by its larger cathode emittance.

Clearly, the simulation in the letter [1] was not performed correctly, and the resultant emittance comparison is misleading. More importantly, the poor emittance performance of the PC scheme, rooted in its physics peculiarity, does not justify its validity as a generic solution to the UE beam problem.

While the PC scheme may have the advantage of technical easiness [6] and achieving high peak brightness without bunch compressor, its physics limitation [2] cannot be circumvented.

**Acknowledgements:** We thank J. Lewellen for insightful comments. This work is supported by the U. S. Department of Energy, Office of Science, Office of Basic Energy Sciences, under Contract No. DE-AC02-06CH11357.